# Theoretical investigations of superconducting MAX phases Ti$_2$InX (X = C, N)

M. Roknuzzaman, A.K.M.A. Islam[*]

*Department of Physics, Rajshahi University, Rajshahi 6205, Bangladesh*

**Abstract**

The structural, elastic, electronic, thermal and optical properties of superconducting MAX phases Ti$_2$InX (X = C, N) are investigated by density functional theory (DFT). The results obtained from the least studied nitride phase are discussed in comparison with those of carbide phase having $T_c$-value half as that of the former. The band structure and density of states show that these phases are conductors, with contribution predominantly from the Ti 3$d$ states. The bulk modulus, Debye temperature, specific heats, thermal expansion coefficient are all obtained as a function of temperature and pressure for the first time through the quasi-harmonic Debye model with phononic effects. Ti$_2$InC and Ti$_2$InN are indicated to be moderately coupled superconductors. The thermal expansion coefficients for both the phases are calculated, and the calculation is in fair agreement with the only available measured value for Ti$_2$InC. Further the first time calculated optical functions reveal that the reflectivity is high in the IR-visible-UV region up to ~ 10 eV and 12.8 eV for Ti$_2$InC and Ti$_2$InN, respectively showing these to be promising coating materials.

## 1. Introduction

The so-called nanolaminates (or MAX) phases since their discovery by Nowotny *et al.* [1] have attracted a lot of interest among the research community due to their remarkable properties having attributes of both ceramic and metal [2-24]. Ceramic attributes include lightweight, elastically rigid, high temperatures strength, whereas metallic attributes show the phases to be thermally and electrically conductive, quasi-ductile and damage tolerant. Currently there are about 60 synthesized MAX phases [3]. Out of these only seven low-$T_c$ superconductors have so far been identified. These are: Mo$_2$GaC [4], Nb$_2$SC [5], Nb$_2$SnC [6], Nb$_2$AsC [7], Ti$_2$InC [8], Nb$_2$InC [9], and Ti$_2$InN [10].

The X-ray diffraction, magnetic and resistivity measurements discovered that Ti$_2$InX (X = C, N) are superconductors [8, 10] with superconducting temperatures of 3.1 and 7.3K, respectively. In fact Bortolozo *et al.* [10] in 2010 showed unambiguously that Ti$_2$InN is the first nitride superconductor belonging to the M$_{n+1}$AX$_n$ family. Among the ternary phases almost all the studies are concerned with carbide properties, but a very limited work on nitrides which was discovered in 1963 by Jeitschko *et al.* [12]. This nitride crystallizes in the same prototype structure as carbides (Cr$_2$AlC), where the basic structural component is an octahedron of six Ti atoms with an N atom instead of C [16]. It has also been shown that the interactions in the Ti$_6$N octahedra are stronger than those in TiN octahedra in agreement with the general trend known for binary carbides and nitrides [16]. Further calculations show that the nitride phase has higher density of states at Fermi level than that of carbide phase. All these point to the role of N atom in changing the electronic structure and the possible transport properties which were the motivation of Bortolozo *et al.* [10] to seek superconductivity in nitride phase. These motivate us to revisit the system Ti$_2$InX (X = C, N) and investigate further the influence of the substitution of N for C on the M$_2$AX nanolaminates.

---

[*] Corresponding author. Tel.: +88 0721 750980; fax: +88 0721 750064.
E-mail address: azi46@ru.ac.bd (A.K.M.A. Islam).

Some works on elastic and electronic structures of Ti$_2$InC have been carried out by several groups of workers [15-20, 22, 23] using several different methodologies. A theoretical study of the elastic properties for six of the seven known superconducting MAX phases: Nb$_2$SC, Nb$_2$SnC, Nb$_2$AsC, Nb$_2$InC, Mo$_2$GaC, and Ti$_2$InC has been presented by Shein *et al.* [20]. Long before this Ivanovskii *et al.* [16] calculated the electronic structure of the H-phases Ti$_2$MC and Ti$_2$MN (M = Al, Ga, In) by the self-consistent linearized muffin-tin-orbital method in the atomic-sphere approximation and the MO LCAO method using RMH parametrization. The band structure and bonding configuration of the H-phases are compared with those of other Ti-M-C and Ti-M-N phases. The energy band structure of the Ti$_2$InC along with some other MAX phases has been calculated in the framework of the full-potential augmented-plane-wave method under GGA [17]. Medkour *et al.* [18] reported on the electronic properties of only M$_2$InC phases by employing the pseudo potential plane wave (PP-PW) method using CASTEP. He *et al.* [19] have performed *ab initio* calculations for the structural, elastic, and electronic properties of only M$_2$InC. Benayad *et al.* [22] very recently included Ti$_2$InN along with Ti$_2$InC to investigate the structural, elastic and electronic properties by using the full-potential linear muffin-tin orbital (FP-LMTO) method. The exchange and correlation potential is treated by the local density approximation (LDA).

Despite all the above efforts, it is clear that T$_2$InN has been subjected to limited study. Moreover full optical as well as finite-temperature and finite-pressure thermodynamical studies are absent for both the superconducting phases. Therefore there is a need to highlight those areas where little or no work has been carried out. We are thus inclined to address these areas of the two nanolaminates as well as revisit the existing theoretical works so as to provide elastic, electronic properties of the carbide phase in comparison with nitride phase. The optical properties such as dielectric function, absorption spectrum, conductivity, energy-loss spectrum and reflectivity for both the phases will be calculated and discussed.

## 2. Computational techniques

The *ab initio* calculations were performed using the plane-wave pseudopotential method within the framework of the density functional theory [25] implemented in the CASTEP code [26]. The ultrasoft pseudopotentials were used in the calculations, and the plane-wave cutoff energy was 500 eV. The exchange-correlation terms used are of the Perdew-Berke-Ernzerhof form of the generalized gradient approximation [27]. We have used a 10×10×2 Monkhorst-pack [28] grid to sample the Brillouin zone. All the structures were relaxed by BFGS methods [29]. Geometry optimization was performed using convergence thresholds of 1×10$^{-5}$ eV/atom for the total energy, 0.03 eV/Å for the maximum force, 0.05GPa for maximum stress, and 1×10$^{-3}$ Å for the maximum displacements. The elastic constants $C_{ij}$, bulk modulus $B$ and electronic properties were directly calculated by the CASTEP code.

The quasi-harmonic Debye model [30] has been employed to investigate the finite-temperature and finite-pressure thermodynamic properties. Here the thermodynamic parameters can be calculated at any temperature and pressure using the DFT calculated *E-V* data at *T* = 0K, *P* = 0 GPa and the Birch-Murnaghan third order EOS [31].

## 3. Results and discussion

### 3.1. Structural and elastic properties

The superconducting MAX phases Ti$_2$InC and Ti$_2$InN possess the hexagonal structure with space group *P*6$_3$/*mmc* (no. 194) as shown in Fig. 1. The unit cell contains two formula units, and the atoms occupy the following Wyckoff positions: the Ti atoms in the position 4$f$ [(1/3, 2/3, $z_M$), (2/3, 1/3, $z_M$+1/2), (2/3, 1/3, − $z_M$), (1/3, 2/3, − $z_M$+1/2)}, the In atoms in the position 2$d$ {(1/3, 2/3, ¾), (2/3, 1/3, 1/4)], and the C atoms (or, N atoms) in the position 2$a$ [(0, 0, 0), (0, 0, 1/2)], where $z_M$ is the internal parameter [2, 32].

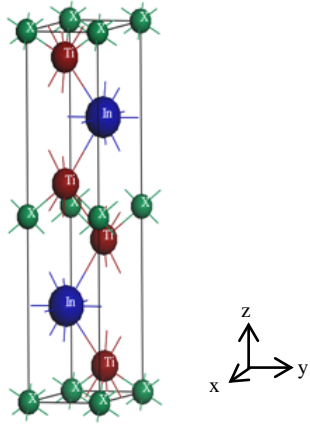

**Fig. 1.** Crystal structure of layered MAX phases Ti$_2$InX (X = C, N).

The calculated fully relaxed equilibrium values of the structural parameters of the two superconducting phases are presented in Table 1 together with other available data on both theoretical [15, 19, 20, 22, 23] and experimental [11, 13, 14]. The comparison shows that the calculated values are in good agreement with the available experimental as well as theoretical results.

**Table 1.** Calculated lattice parameters ($a$ and $c$ in Å), ratio $c/a$ and internal parameters $z_M$ for the superconducting MAX phases Ti$_2$InC and Ti$_2$InN.

| Phase | $a$ | $c$ | $c/a$ | $z_M$ | Ref. |
|---|---|---|---|---|---|
| Ti$_2$InC | 3.1453 | 14.215 | 4.519 | 0.0780 | Present |
|  | 3.1373 | 14.1812 | 4.520 | 0.0783 | [15] |
|  | 3.14 | 14.17 | 4.51 | 0.0779 | [19] |
|  | 3.1485 | 14.2071 | 4.512 | 0.0780 | [20] |
|  | 3.084 | 13.906 | 4.508 | 0.0788 | [22] |
|  | 3.135 | 14.182 | 4.524 |  | [23] |
|  | 3.134 | 14.077 | 4.492 |  | [11][Exptl.] |
|  | 3.133 | 14.10 | 4.5 |  | [14][Exptl.] |
| Ti$_2$InN | 3.0956 | 14.063 | 4.543 | 0.07855 | Present |
|  | 3.033 | 13.727 | 4.525 | 0.07908 | [22] |
|  | 3.07 | 13.97 | 4.54 |  | [13][Exptl.] |

**Table 2** Calculated elastic constants ($C_{ij}$, in GPa), bulk moduli (*B*, in GPa), shear moduli (*G*, in GPa), Young's moduli (*Y*, in GPa), Poisson's ratio (*ν*), *A* and $k_c/k_a$ for superconducting Ti$_2$InC and Ti$_2$InN.

| Phase | $C_{11}$ | $C_{12}$ | $C_{13}$ | $C_{33}$ | $C_{44}$ | *B* | *G* | *Y* | *ν* | *A* | $k_c/k_a$ | Ref. |
|---|---|---|---|---|---|---|---|---|---|---|---|---|
| Ti$_2$InC | 284.2 | 58.7 | 52.3 | 246.1 | 90.0 | 126.4 | 100.4 | 240 | 0.184 | 0.798 | 1.230 | Present |
| | 273.4 | 62.9 | 50.3 | 232.3 | 87.2 | 120 | 96 | 228 | 0.184 | 0.829 | 1.293 | [15] |
| | 282.6 | 70.2 | 54.9 | 232.9 | 57.6 | 124.7 | 81.7 | 201.1 | 0.232 | 0.542 | 1.365 | [20] |
| | 281.0 | 57.7 | 44.5 | 226.6 | 85.8 | | 98.6 | | | 0.768 | 1.371 | [22] |
| | 287.0 | 65 | 53 | 244 | 85 | 128 | 99 | | | 0.766 | 1.288 | [23] |
| Ti$_2$InN | 213.7 | 36.8 | 105.6 | 231.7 | 98 | 125.5 | 81 | 200 | 0.234 | 1.11 | 0.312 | Present |
| | 102.9 | 60.9 | 62.7 | 106.1 | 46.1 | 41.8[a] | 32.9 | 86.3 | 0.31 | 2.19 | 0.884 | [22] |

[a] Calculated based on data from [22].

The elastic constant tensors of the superconducting MAX phases Ti$_2$InC and Ti$_2$InN are reported in Table 2 along with available computed elastic constants [15, 20, 22, 23]. For Ti$_2$InC the agreement with available theoretical results is quite good. But for Ti$_2$InN, the only set of data due to Benayad *et al.* [22] deviate much from our calculations and also from the trend for similar phase (Table 2). The reason may be the use of FP-LMTO method treated with LDA with P-W parameterization.

Using the second order elastic constants, the bulk modulus *B*, shear modulus *G* (all in GPa), Young's modulus *Y*, and Poisson's ratio *ν* at zero pressure are calculated and presented in Table 2. The pressure dependence of the elastic constants is a very important characterization of the crystals with varying pressure and/or temperature, but we defer it till in a later section. The ductility of a material can be roughly estimated by the ability of performing shear deformation, such as the value of shear-modulus-to-bulk-modulus ratios. Thus a ductile plastic solid would show low *G/B* ratio (< 0.5); otherwise, the material is brittle. As is evident from Table 2, the calculated *G/B* ratios are 0.8 and 0.65 for carbide and nitride phases, respectively indicating that first compound is brittle in nature and the second one will be more on the brittle/ductile border line. The same can be inferred from an additional argument that the variation in the brittle/ductile behavior follows from the calculated Poisson's ratio. For brittle material the value is small enough, whereas for ductile metallic materials *ν* is typically 0.33 [24].

The elastic anisotropy of the shear of hexagonal crystals, defined by $A = 2C_{44}/(C_{11}-C_{12})$, may be responsible for the development of microcracks in the material [33]. This factor is unity for an ideally isotropic crystal. The calculated value of *A* increases from 0.798 to 1.11 as C atom is replaced by N. We can also examine a second anisotropy parameter which is the ratio between the uniaxial compression values along the *c* and *a* axis for a hexagonal crystal: $k_c/k_a = (C_{11} + C_{12} - 2C_{13})/(C_{33} - C_{13})$. We find that the compressibility of Ti$_2$InC along the *c*-axis is larger than along the *a*-axis ($k_c/k_a = 1.23$) in agreement with other calculations [15, 20, 22, 23], but for Ti$_2$InN the situation is reversed, as *c* is stiffer for this material.

*3.2. Electronic band structure and bonding*

The energy bands of the two nanolaminates along the high symmetry directions in the first Brillouin zone are shown in Fig. 2 (a,b) in the energy range from −15 to +5 eV. The band structures of both the superconducting phases reveal 2D-like behavior with smaller energy dispersion along the c-axis and in the K–H and L–M directions. The occupied valence bands of Ti$_2$InC and Ti$_2$InN lie in the energy range from − 8.8 eV to Fermi level and − 9.5 eV to Fermi level, respectively. Further, the valence and conduction bands are seen to overlap, thus indicating metallic-like behavior of both the phases. This conductivity increases as C is replaced by N. The In 4$d$ and C 2$s$-type quasi-core bands with a small dispersion can be seen in the energy intervals ∼ −13.7 to −14.4 eV and from –11 to –10 eV, respectively below the Fermi level. The corresponding energy intervals are about -14 to -15 eV for In 4$d$ and N 2$s$-type quasi-core bands. As seen in ref. [15] the multiband character of the systems can be inferred from three near Fermi bands which intersect the Fermi level.

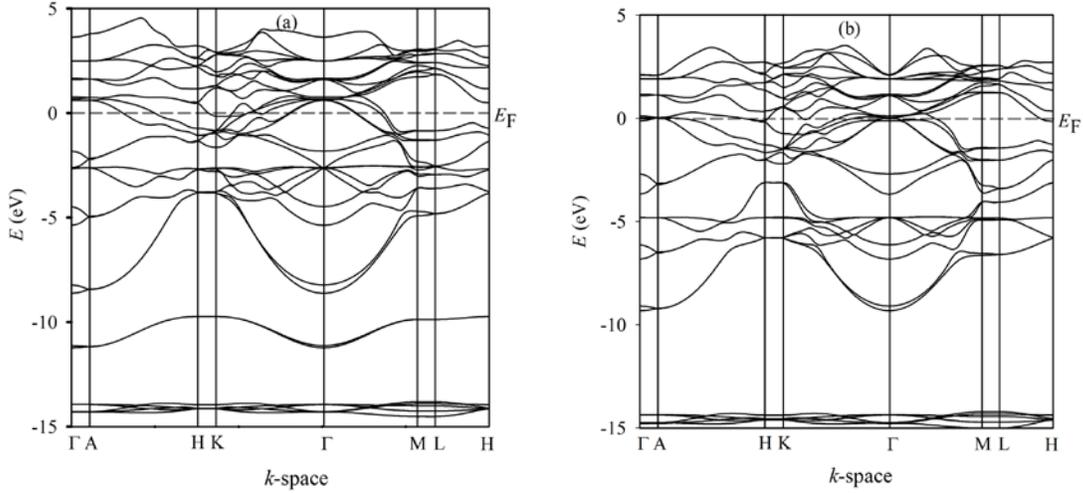

**Fig. 2.** Calculated band structures of (a) Ti$_2$InC and (b) Ti$_2$InN.

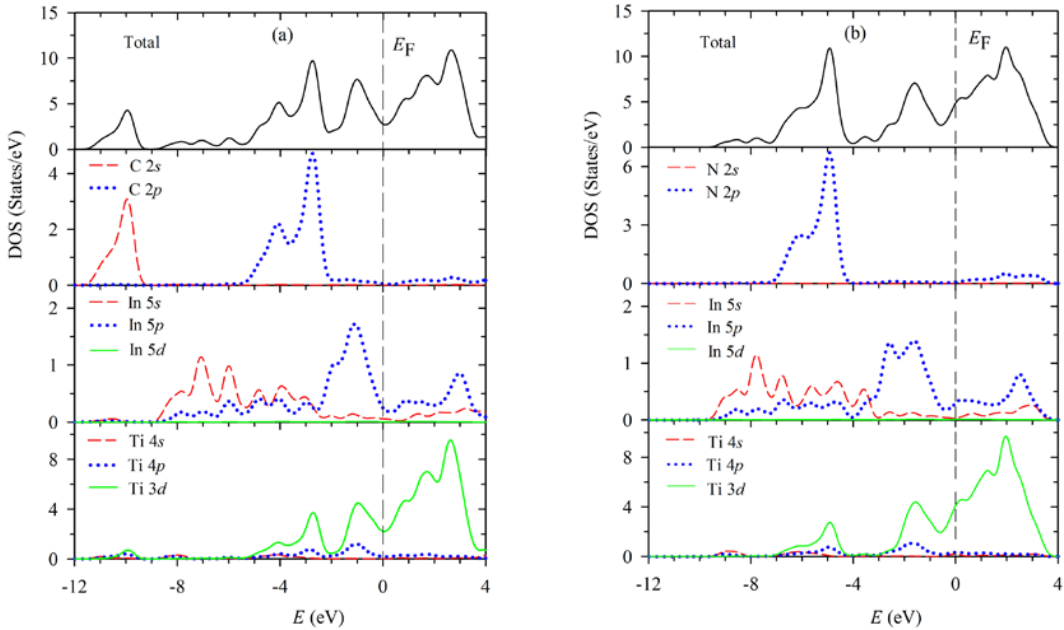

**Fig. 3.** Total and partial DOSs of (a) Ti$_2$InC and (b) Ti$_2$InN.

The total and partial densities of states for the two superconducting phases are illustrated in Fig. 3 (a, b). The values of DOS at the Fermi level are 2.78 and 4.98 states/eV which predominantly contain contributions from the Ti 3$d$ states of 2.22 and 4.06 states/eV of the two phases $Ti_2InC$ and $Ti_2InN$, respectively. The diffuse character of both $s$ and $p$ states of In atoms causes larger dispersion of In bands than those due to C and N. A covalent interaction occurs (− 9 eV to Fermi level) between the constituting elements as a result of the degeneracy of the states with respect to both angular momentum and lattice site. C $p$, N $p$, and Ti $d$ as well as In $p$ and Ti $d$ states are all hybridized. Such hybridization peak of Ti $d$−C $p$ in $Ti_2InC$ and Ti $d$−N $p$ in $Ti_2InN$ lies lower in energy (−5 to −2 eV) and (−7 to −4 eV) than that of Ti $d$−In $p$ (−3 eV to Fermi level). All these indicate that Ti-In bond is weaker than either Ti–C or Ti–N bond. The population analysis shows that bond lengths in Å for $Ti_2InC$ and $Ti_2InN$ in increasing order are as: Ti-C (2.1277), Ti-Ti (2.8661), Ti-In (3.0456), In-C (3.9908) and Ti-N (2.1010), Ti-Ti (2.8416), Ti-In (3.0013), In-N (3.9439). The bands associated with N atoms are narrower and lower in energy. This is attributed to the large electronegativity of N compared to that of C.

Ivanovskii *et al.* [16] from their band structure calculations for the phases suggest that the transition metal does not play role in the superconducting mechanism suggesting that the transport behavior of this material is of 2-D nature. The C atom is less electronegative than N, and the chemical bond between Ti-C is less polarized than Ti-N. It is thus hypothesized [16] that the electrons of the basal plane rather than the $d$-electrons of Ti may be responsible for the superconducting behavior in nanolaminates. One also notes that $T_c$ value is more than doubled when C atoms are replaced by N atoms in the $Ti_2InX$ compound.

*3.3. Thermodynamic properties at elevated temperature and pressure*

The elastic parameters and associated physical quantities like Debye temperature etc. allow a deeper understanding of the relationship between the mechanical properties and the electronic and phonon structure of materials. We investigated the thermodynamic properties of $Ti_2InC$ and $Ti_2InN$ by using the quasi-harmonic Debye model, the detailed description of which can be found in literature [30]. For this we need $E$-$V$ data obtained from Birch-Murnaghan third order EOS [31] using zero temperature and zero pressure equilibrium values, $E_0$, $V_0$, $B_0$, based on DFT method. Then the thermodynamic properties at finite-temperature and finite-pressure can be obtained using the model. The non-equilibrium Gibbs function $G^*(V; P, T)$ can be written in the form [30]:

$$G^*(V;P,T) = E(V) + PV + A_{vib}[\Theta(V);T] \qquad (1)$$

where $E(V)$ is the total energy per unit cell, $PV$ corresponds to the constant hydrostatic pressure condition, $\Theta(V)$ is the Debye temperature, and $A_{vib}$ is the vibrational term, which can be written using the Debye model of the phonon density of states as [30]:

$$A_{vib}(\Theta,T) = nkT\left[\frac{9\Theta}{8T} + 3\ln(1-\exp(-\Theta/T)) - D\left(\frac{\Theta}{T}\right)\right] \qquad (2)$$

where $n$ is the number of atoms per formula unit, $D(\Theta/T)$ represents the Debye integral.

A minimization of $G^*(V; P, T)$ with respect to volume $V$ can now be made to obtain the thermal equation of state $V(P, T)$ and the chemical potential $G(P, T)$ of the corresponding phase. Other macroscopic properties can also be derived as a function of $P$ and $T$ from standard thermodynamic relations [30]. Here we computed the bulk modulus, Debye temperature specific heats, and volume thermal expansion coefficient at different temperatures and pressures.

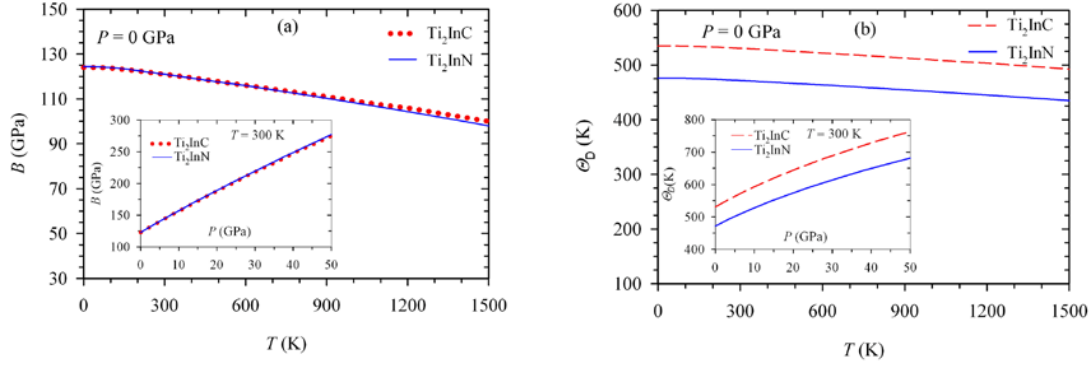

**Fig. 4.** Temperature dependennce of (a) Bulk modulus and (b) Debye temperature of $Ti_2InC$ and $Ti_2InN$. *Inset* shows pressure variation.

The temperature variation of isothermal bulk modulus $B$ of $Ti_2InC$ and $Ti_2InN$ is shown in Fig. 4 and the *inset* of which shows $B$ as a function of pressure. We see that there is hardly any difference in the values of $B$ for the two phases and these vary identically as a function of temperature. Furthermore, it is found that the bulk modulus increases with pressure at a given temperature and decreases with temperature at a given pressure, which is consistent with the trend of volume.

Fig. 4 displays temperature dependence of Debye temperature $\Theta_D$ at zero pressure of $Ti_2InC$ and $Ti_2InN$. One observes that $\Theta_D$, smaller for nitride phase, decrease non-linearly with temperature for both the phases. Further $\Theta_D$ presented as *inset* of Fig. 4 (a) at $T = 300K$ shows a non-linear increase. The variation of $\Theta_D$ with pressure and temperature reveals that the thermal vibration frequency of atoms in the nanolaminates changes with pressure and temperature. We can estimate the value of the electron-phonon coupling constant ($\lambda$) can be estimated from McMillan's relation [34] using the calculated Debye temperature and the measured $T_c$. With a typical value of Coulomb repulsion constant ($\mu^* = 0.13$), we find $\lambda \sim 0.49$, and 0.62, for $Ti_2InC$ and $Ti_2InN$, respectively. The values imply that both of these are moderately coupled superconductors.

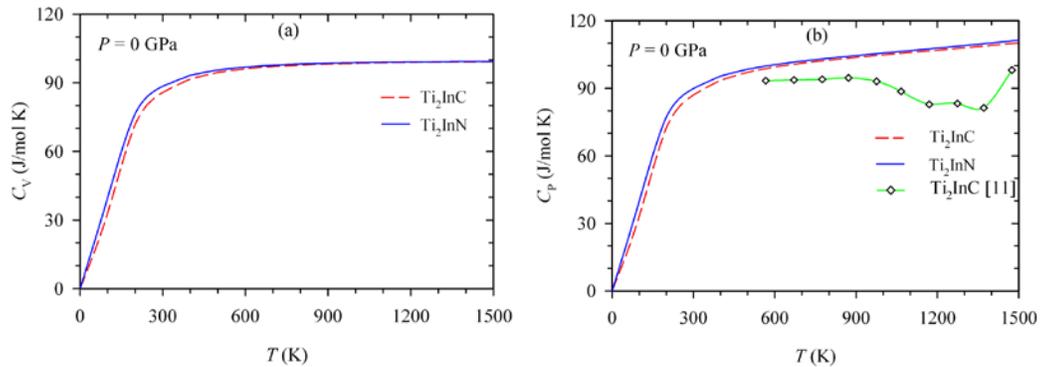

**Fig. 5.** Temperature dependence of (a) specific heat at constant pressure, and (b) specific heat at constant volume of $Ti_2InC$ and $Ti_2InN$.

Fig. 5 (a, b) show the temperature dependence of constant-volume and constant-pressure specific heat capacities $C_V$, $C_P$ of $Ti_2InC$ and $Ti_2InN$. We know that phonon thermal softening occurs when the temperature increases and hence the heat capacities increase with increasing temperature. It should be noted that the heat capacity anomaly close to $T_c$-value (3.1 and 7.3 K, for the two superconductors) is so small (about 0.1%) that it has no effect on the analysis being made here. The only measured $C_P$ data for Ti$_2$InC due to Barsoum *et al.* [11] show complex behavior as shown on the theoretical graph. Even the authors themselves remark that such a complex behavior is not expected from a single phase solid that does not go through phase transitions. The drop in $C_p$ must be related to loss of In atoms from the sample. This type of loss would be endothermic and thus exhibits a trough as observed. Barsoum *et al.* [11] acknowledged that the heat capacity measurements should be repeated with larger samples where the surface to volume ratio is reduced. The increase at higher temperatures is most likely due to oxidation.

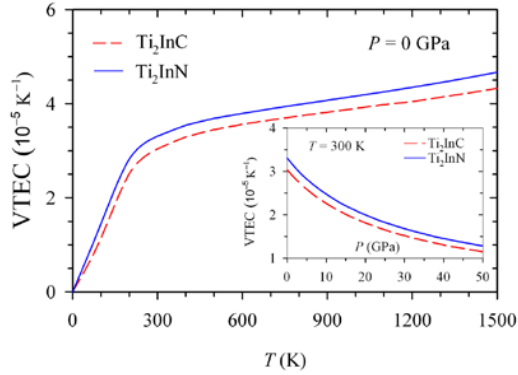

**Fig. 6.** Temperature dependent thermal expansion co-efficient of Ti$_2$InC and Ti$_2$InN. *Inset* shows pressure variation.

The volume thermal expansion coefficient (VTEC), $\alpha_V$ as a function of both temperature and pressure is displayed in Fig. 6. The expansion coefficient is seen to increase rapidly especially at temperature below 300K, whereas it gradually tends to a slow increase at higher temperatures. On the other hand at a constant temperature, the expansion coefficient decreases strongly with pressure. It is well-known that the thermal expansion coefficient is inversely related to the bulk modulus of a material. The calculated values of $\alpha_V$ at 300 K for Ti$_2$InC and Ti$_2$InN are $3.04\times10^{-5}$ and $3.3\times10^{-5}$ K$^{-1}$, respectively. The measured value of linear thermal expansion coefficient of Ti$_2$InC is $9.5\times10^{-6}$K$^{-1}$ [11]. Assuming, linear thermal expansion coefficient = $\alpha_V/3$, the calculated value of $10.1\times10^{-6}$K$^{-1}$ for Ti$_2$InC is in fair agreement with experiment.

*3.3. Optical properties*

The study of the optical functions of solids provides a better understanding of the electronic structure. The imaginary part of complex dielectric function, $\varepsilon(\omega) = \varepsilon_1(\omega) + i\varepsilon_2(\omega)$, is obtained from the momentum matrix elements between the occupied and the unoccupied electronic states. This is calculated directly using [35]:

$$\varepsilon_2(\omega) = \frac{2e^2\pi}{\Omega\varepsilon_0} \sum_{k,v,c} \left|\psi_k^c|\mathbf{u}\cdot\mathbf{r}|\psi_k^v\right|^2 \delta\left(E_k^c - E_k^v - E\right) \qquad (3)$$

where $\psi_k^c$ and $\psi_k^v$ are the conduction and valence band wave functions at $k$, respectively, $\mathbf{u}$ is the vector defining the polarization of the incident electric field, $\omega$ is the light frequency and $e$ is the electronic charge and. The Kramers-Kronig transform of the imaginary part $\varepsilon_2(\omega)$ provides the real part. Eqs. 49 to 54 in ref. [35] define all other optical constants, such as refractive index, absorption spectrum, loss-function, reflectivity and conductivity (real part).

The calculated optical functions of $Ti_2InC$ and $Ti_2InN$ for photon energies up to 20 eV for polarization vectors [100] and [001] (only spectra for [100] shown) along with measured spectra of TiC and TiN (where available) are shown in Fig 7. We have used a 0.5 eV Gaussian smearing for all calculations. The calculations only include interband exciatations. In metal and metal-like systems there are intraband contributions from the conduction electrons mainly in the low-energy infrared part of the spectra. It is thus necessary to include this via an empirical Drude term to the dielectric function [36, 37]. A Drude term with plasma frequency 3 eV and damping (relaxation energy) 0.05 eV was used.

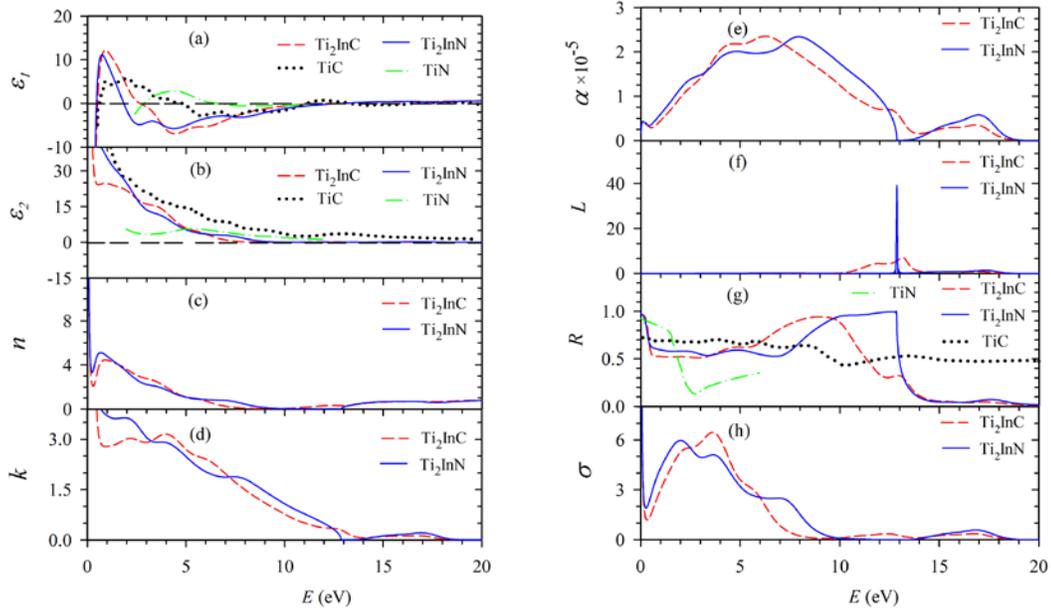

**Fig. 7.** Energy dependent (a) real part of dielectric function, (b) imaginary part of dielectric function, (c) refractive index, (d) extinction coefficient, (e) absorption, (f) loss function, (g) reflectivity and (h) real part of conductivity of $Ti_2InC$ and $Ti_2InN$ along [100] direction. Experimental data shown for TiC and TiN are from ref. [38] and [39], respectively.

Despite some variation in heights and positions of peaks, the overall features of our calculated optical spectra of $Ti_2InC$ and $Ti_2InN$ are roughly similar. In the energy range for which $\varepsilon_1(\omega) < 0$, $Ti_2InC$ and $Ti_2InN$ exhibit the metallic characteristics (Fig. 7 (a)). The result of $Ti_2InC$ is somewhat different as regards the energy range for negativity of $\varepsilon_1(\omega)$. Both the supeconducting nanolaminates have positive static dielectric constant $\varepsilon_1(0)$.. The dielectric function of $Ti_2InC$ is compared with that of $TiC_{0.9}$ [38]. We see that the double peak structure centered at 1.7 eV for $TiC_{0.9}$ is replaced with a sharp peak at around 0.7 eV for $Ti_2InC$. The spectra differ at low energy due to the electronic structure change near the Fermi level, induced by the addition of In layer in TiC. The same inference can be made when one compares low energy spectra of $Ti_2InN$ and TiN [39]. On the other hand no maxima are seen in $\varepsilon_2$ for the two MAX phases, although the values are large in the low energy region (Fig. 7 (b)). The corresponding spectra for $TiC_{0.9}$ [38] and TiN [39] are shown for comparison. The refractive index and extinction coefficients of the nanolaminates are displayed in Fig. 7 (c) and (d).

The absorption coefficient provides data about optimum solar energy conversion efficiency and it indicates how far light of a specific energy (wavelength) can penetrate into the material before being absorbed. Fig. 7 (e) shows the absorption coefficients of both the phases which begin at 0 eV due to their metallic nature. $Ti_2InC$ has two peaks, one at ~ 4.3 eV (same for $Ti_2InN$) and the other at 6.3 eV (8 eV for $Ti_2InN$), besides having a shoulder at lower energy. Both the nanolaminates show rather good absorption coefficient in the 4 – 10 eV region. The energy loss $L(\omega)$ of a fast electron traversing in the material is depicted in Fig. 7 (f). The bulk plasma frequency $\omega_P$ is at the peak position which occurs at $\varepsilon_2 < 1$ and $\varepsilon_1 = 0$. In the energy-loss spectrum, we see that $\omega_P$ of the two phases $Ti_2InC$ and $Ti_2InN$ are ~13.2 eV and ~ 12.8 eV, respectively. When the incident photon frequency is higher than $\omega_P$, the material becomes transparent.

Fig. 7 (g) presents the reflectivity spectra as a function of photon energy in comparison with measured spectra of $TiC_{0.97}$ [38] and TiN [40]. The reflectance for $TiC_{0.97}$ is nearly constant over the energy range considered. With addition of In to TiC the reflectivity is much higher in the infrared region, it then decreases sharply to 0.55 which becomes almost steady till 5 eV. After an increase with photon energy up to ~ 10 eV, the reflectivity falls again. On the other hand we find that the reflectivity in $Ti_2InN$ is high in IR-visible-UV up to ~12.8 eV region (reaching maximum between 10 and 12.8 eV). Compared to this the reflectivity of TiN [40] starts with a higher value in the infrared and there is a sharp drop between 2 and 3 eV, which is characteristics of high conductance. The low reflectivity in the region of visible blue and violet light (2.8-3.5 eV) increases to a value of 0.36 at 6 eV (ultraviolet). The analysis shows that the nitride phase would be a comparatively better material as promising candidate for use as coating material.

Fig. 7 (h) shows that the photoconductivity starts with zero photon energy due to the reason that the materials have no band gap which is evident from band structure. Moreover, the photoconductivity and hence electrical conductivity of a material increases as a result of absorbing photons.

## 4. Conclusion

We have performed a first-principles calculations based on DFT to compare the structural, elastic, thermodynamic, electronic and optical properties of the two superconducting MAX phases $Ti_2InC$ and $Ti_2InN$. The obtained elastic constants are compared with available calculations and elastic anisotropy discussed. The carbide phase is found to be brittle in nature, while the nitride phase is less brittle (near the border line).

The energy band structure and total densities of states reveal that both the materials exhibit metallic conductivity. This conductivity increases as X is changed from C atom to N in $Ti_2InX$. Hybridization of Ti-atom $d$ states with C (N)-atom $p$ states is responsible for the bonding. The Ti-In bond is weaker and the order of the bond strength: Ti–N > Ti–C > Ti-In. The bands associated with N atoms are lower in energy and narrower that can be attributed to the large electronegativity of N compared to that of C.

The temperature and pressure dependence of bulk modulus, specific heats, thermal expansion coefficient, and Debye temperature are all obtained through the quasi-harmonic Debye model, and the results are analyzed. The estimated electron-phonon coupling strengths $\lambda$ ~ 0.49, and 0.62, for $Ti_2InC$ and $Ti_2InN$, respectively imply that both are moderately coupled superconductors. The heat capacities increase with increasing temperature, which shows that phonon thermal softening occurs when the temperature increases. The thermal expansion coefficients for $Ti_2InC$ and $Ti_2InN$ are evaluated, and the calculation is in fair agreement with the only measured value available for $Ti_2InC$.

The optical features such as the real and imaginary parts of the dielectric function and positive dielectric constant do show to support the potential applications of the compounds in future. The reflectivity is high in the IR-visible-UV region up to ~ 10 eV and 12.8 eV for $Ti_2InC$ and $Ti_2InN$, respectively showing promise as good coating materials.

**References**

[1]  H. Nowotny, Prog. Solid State Chem. 2 (1970) 27.
[2]  M.W. Barsoum, Prog. Solid State Chem. 28 (2000) 201.
[3]  P. Eklund, M. Beckers, U. Jansson, H. Högberg, and L. Hultman, Thin Solid Films 518 (2010) 1851.
[4]  L. E. Toth, J. Less Common Met. 13, 129 (1967).
[5]  K. Sakamaki, H. Wada, H. Nozaki, Y. Onuki, and M. Kawai, Solid State Commun. 112, 323 (1999).
[6]  A. D. Bortolozo, O. H. Sant'Anna, M. S. da Luz, C. A. M. dos Santos, A. S. Pereira, K. S. Trentin, and A. J. S. Machado, Solid State Commun. 139, 57 (2006).
[7]  S. E. Lofland, J. D. Hettinger, T. Meehan, A. Bryan, P. Finkel, S. Gupta, M. W. Barsoum, and G. Hug, Phys. Rev. B 74, 174501 (2006).
[8]  A. D. Bortolozo, O. H. Sant'Anna, C. A. M. dos Santos, and A. J. S. Machado, Solid State Commun. 144, 419 (2007).
[9]  A. D. Bortolozo, Z. Fisk, O. H. Sant'Anna, C. A. M. dos Santos, and A. J. S. Machado, Physica C 469, 256 (2009).
[10] A. D. Bortolozo, G. Serrano, A. Serquis, D. Rodrigues Jr., C. A. M. dos Santos, and Z. Fisk, Solid State Communications 150 (2010) 1364-1366.
[11] M. W. Barsoum, J. Golczewski, H. J. Seifert and F. Aldinger, J. Alloys Compd. 340 (2002) 173.
[12] W. Jeitschko, H. Nowotny and F. Benesovsky, Monatsh. Chem. 94 (1963) 1201.
[13] W. Jeitschko, H. Nowotny and F. Benesovsky, Monatsh. Chem. 95 (1964) 178.
[14] B. Manoun, O. D. Leaffer, S. Gupta, E. N. Hoffman, S. K. Saxena, J. E. Spanier and M. W. Barsoum, Solid State Commun. 194 (2009) 1978.
[15] I.R. Shein, A.L. Ivanovskii, Phys. of the Solid State, 51 (2009) 1608-1612.
[16] A. L. Ivanovskii, R. F. Sabiryanov, A. N. Skazkin, V. M. Zhukovskii and G. P. Shveikin, Inorg. Mater. 36 (2000) 28.
[17] G. Hug, Phys. Rev. B 74 (2006) 184113.
[18] Y. Medkour, A. Bouhemadou, A. Roumili, Solid State Communications 148 (2008) 459-463.
[19] X. He, Y. Bai, Y. Li, C. Zhu, M. Li. Solid State Commun. 149 (2009) 564-566.
[20] I. R. Shein and A.L. Ivanovskii, Phys. Status Solidi B 248 (1) (2011) 228–232.
[21] O.D. Leaffer, S. Gupta, M.W. Barsoum, and J.E. Spaniera, J. Mater. Res. 22 (2007) 2651-2654.
[22] N. Benayad, D. Rached, R. Khenata, F. Litimein, A.H. Reshak, M. Rabah, H. Baltache, Modern Phys. Lett. B, 25 (2011) 747-761.
[23] B. Liu, J.Y. Wang, J. Zhang, J.M. Wang, F.Z. Li, Y.C. Zhou. Appl. Phys. Lett. 94 (2009) 181906.
[24] J. Haines, J. M. Leger, and G. Bocquillon, Annu. Rev. Mater.Res. 31, 1 (2001).
[25] W. Kohn, L.J. Sham, Phys. Rev. A 140 (1965) 1133.
[26] S.J. Clark, M.D. Segall, C.J. Pickard, P.J. Hasnip, M.J. Probert, K. Refson, M.C. Payne, Zeitschrift fuer Kristallographie 220 (2005) 567.
[27] J.P. Perdew, K. Burke, M. Ernzerhof, Phys. Rev. Lett. 77 (1996) 3865.
[28] H.J. Monkhorst, J.B. Pack, Phys. Rev. B 13 (1976) 5188.
[29] T.H. Fischer, J. Almlof, J. Phys. Chem. 96 (1992) 768.
[30] M.A. Blanco, E. Francisco, V. Luaña, Comput. Phys. Comm. 158 (2004) 57–72.
[31] F. Birch, J. Geophy. Res. 83 (1978) 1257.
[32] M.W. Barsoum, in *Encyclopedia of Materials: Science and Technology* (Elsevier, Amsterdam, 2006) p. 1.

[33] N.I. Medvedeva, A. N. Enyashin, and A. L. Ivanovskii, J. Structural Chem. 52 (4) (2011) 785-802.
[34] W. L. McMillan, Phys. Rev. B 167 (1967) 331.
[35] Materials Studio CASTEP manual © Accelrys 2010.
http://www.tcm.phy.cam.ac.uk/castep/documentation/WebHelp/CASTEP.html
[36] S. Li, R. Ahuja, M. W. Barsoum, P. Jena, and B. Johansson, Appl. Phys. Lett. 92 (2008) 221907.
[37] R. Saniz, L. Ye, T. Shishidou, and A. J. Freeman, Phys. Rev. B 74 (2006) 014209.
[38] D. W. Lynch, C. G. Olson, D. J. Peterman, J. H. Weaver,Phy. Rev. B 22(1980) 3991-3997
[39] R. Eibler, M. Dorrer, and A. Neckel, J. Phys. C 16 (1983) 3137.
[40] B. Karlsson, "Optical properties of solids for solar energy conversion," Ph.D. thesis, Acta Universitatis Upsaliensis, Uppsala, Sweden, 1981.